\documentclass[a4paper,fleqn,usenatbib]{mnras}

\usepackage[T1]{fontenc}
\usepackage{ae,aecompl}
\usepackage[varg]{txfonts}
\usepackage{graphicx}
\usepackage{caption}
\usepackage{aas_macros}
\usepackage{graphicx}	% Including figure files

\usepackage{savesym}
\savesymbol{iint}
\usepackage{wasysym}
\restoresymbol{WAS}{iint}

\title[Bright X-Ray Flares]{Bright X-ray flares from Sgr~A*}

\author[G.D. Karssen et al.]{
G. D. Karssen$^{1}$,
M. Bursa$^{2}$,
A. Eckart$^{1,3}$,
M. Valencia-S.$^{1}$,
M. Dov\u{c}iak$^{2}$,
V. Karas$^{2}$,
J. Hor\'ak$^{2}$
\\
$^{1}$I. Physikalisches Institut, Universit\"{a}t zu K\"{o}ln,
Z\"{u}lpicher Str. 77, 50937 K\"{o}ln, Germany \\
$^{2}$Astronomical Institute, Academy of Sciences,
Bo\u{c}n\'{i} II 1401, CZ-14131 Prague, Czech Republic\\
$^{3}$Max-Planck-Institut f\"{u}r Radioastronomie,
  Auf dem H\"{u}gel 69, 53121 Bonn, Germany
}

\date{Accepted 2017 September 4. Received 2017 September 4; in original form 2017 May 22}

\pubyear{2015}

\begin{document}
\label{firstpage}
\pagerange{\pageref{firstpage}--\pageref{lastpage}}
\maketitle

\begin{abstract}
We address a question whether the observed light curves of X-ray flares 
originating deep in galactic cores can give us independent constraints on the mass 
of the central supermassive black hole. To this end we study four brightest flares 
that have been recorded from Sagittarius A*. They all exhibit an asymmetric shape 
consistent with a combination of two intrinsically separate peaks that occur at a 
certain time-delay with respect to each other, and are characterized by their mutual 
flux ratio and the profile of raising/declining parts. Such asymmetric shapes arise 
naturally in the scenario of a temporary flash from a source orbiting near a supermassive 
black hole, at radius of only $\sim10$--$20$ gravitational radii. An interplay of 
relativistic effects is responsible for the modulation of the observed light curves: 
Doppler boosting, gravitational redshift, light focusing, and light-travel time delays. 
We find the flare properties to be in agreement with the simulations 
(our ray-tracing code {\tt sim5lib}). The inferred mass for each of the flares comes 
out in agreement with previous estimates based on orbits of stars; the latter have been 
observed at radii and over time-scales two orders of magnitude larger than those 
typical for the X-ray flares, so the two methods are genuinely different. 
We test the reliability of the method by applying it to another object, namely, 
the Seyfert~I galaxy RE~J1034+396.
\end{abstract}


%%%%%%%%%%%%%%%%%%%%%%%%%%%%%%%%%%%%%%%%%%%%%%%%%%

%%%%%%%%%%%%%%%%% BODY OF PAPER %%%%%%%%%%%%%%%%%%

\section{Introduction}
The radio source Sagittarius~A* (Sgr~A*) at the dynamical center of our galaxy
\citep{Balick1974} shows variability in all wavelengths visible to us 
\citep[e.g.][]{Brown1982,Macquart2006,Eckart2004,Eckart2006,Eckart2008,Hornstein2007,Yusef-Zadeh2006,Yusef-Zadeh2009,Dodds-Eden2009}.
Sgr~A* is usually visible at sub-mm and radio frequencies, while 
detectable flares at near infrared (NIR) wavelengths can be observed about 2-6 
times per day \citep{Genzel2003,Ghez2004,Yusef-Zadeh2006,Nishiyama2009}.
The biggest
excursions have been observed at X-ray frequencies, although flares at
near-infrared frequencies occur more often: X-ray flares about 10 time brighter than the
quiescent background flux occur about once per day \citep{Baganoff2001}.
The first observation of an
X-ray flare of Sgr~A* was reported by \citet{Baganoff2001} and shows a peculiar
asymmetry, it consists of two distinct peaks and a drop
in between. The authors attribute the emission to accretion processes near
the super-massive black hole (SMBH)
\citep[for a recent review on the validity of that assumption see][]{Eckart2017}.
  \citet{Porquet2003} published an
observation of an even brighter flare, also with an asymmetric shape consisting
of two peaks. Another bright flare was observed by \citet{Porquet2008}, together
with some less luminous ones, which will not be discussed further in this
treatise. The bright flare however, shows again an asymmetrical shape. One of the
latest very bright X-ray flare was published by \citet{Nowak2012} and it also
exhibits an asymmetric shape, which is modeled by the authors as two Gaussians.
Here, only these four brightest X-ray flares will be analyzed, since it is not
feasible to make accurate statements about less bright flares, due to their limited
signal to noise ratio.

Hotspot models \citep{Stella1998,Stella1999} have been tested as a mechanism to explain the flares of Sgr~A*
by numerous authors, mainly applied to near infrared observations
\citep{Genzel2003,Broderick2005,Broderick2006,Meyer2006a,Meyer2006,Meyer2007,Trippe2007,Dexter2009,Zamaninasab2010,Zamaninasab2011}
to look for quasi periodic oscillations (QPO) appearing as substructures of some
NIR-flares that were observed from Sgr~A*.
\citet{Eckart2004} reported the first simultaneous observations of a particular Sgr~A* flare
at near infrared as well as X-ray wavelengths. This suggests that NIR and X-ray
flares of Sgr~A* have the same origin and result from the same physical process.
\citet{Mossoux2014} have applied a hotspot model to an X-ray flare of Sgr~A*, but
find that this scenario is not sufficient to model this particular flare.

In the following we will use the term 'hotspot' or blob if we refer to 
relativistic orbital motion of a luminous matter component around the 
supermassive black hole as the prime reason for the observed phenomena.
Here, we argue that the this scenario is a good model for Sgr~A*'s X-ray 
variability. Furthermore, this scenario makes it possible to constrain the mass of the
central black hole.
There have already been a number of different attempts to constrain the mass which must be
located at the position of Sgr~A*.
\citet{Wollman1977} applied a virial analysis to the radial velocities of ionized
gas and determined a mass of $ (2-4) \times 10^6 M_{\odot}$. However, gas
is not just influenced by gravity, and so measurements of the velocity dispersion
of stars is a better estimate for the mass which is, however, in agreement with those
resulting from gas motions \citep{Rieke1988,McGinn1989,Sellgren1990,Haller1996}.
Due to the large radius of this method, within which the mass must be enclosed, 
it is not possible to infer the existence of a black hole from these measurements
alone.
Employing the use of virial estimators and estimators of \citet{Bahcall1981},
\citet{Genzel1996} arrived at a central mass of $3 \times 10^6 M_{\odot}$, 
surrounded by a star cluster with a mass of $10^6 M_{\odot}$.
\citet{Eckart2002} showed for the first time that S-cluster star S2 is on a 
bound orbit around SgrA*.  Improved orbital elements of S2 were given shortly 
after by \citet{Schoedel2002} and \citet{Ghez2003}. Using Kepler's third law, the 
enclosed mass can be estimated, if the distance to the Galactic center can be 
accurately determined. \citet{Gillessen2009} analyzed the orbits of several 
stars of the S-cluster and found an enclosed mass of $4.3 \times 10^6 M_{\odot}$.
The latest measurements of Sgr~A*'s mass and its distance to us employed multiple
stellar orbits and were published by \citet{Boehle2016}. The authors find a mass
estimate of $4.02 \pm 0.16 \pm 0.04 \times 10^6 M_{\odot}$ and their estimate on
the distance is $7.86 \pm 0.14 \pm 0.04$ kpc. The two uncertainties correspond to the 
statistical and the systematic part of the uncertainties. Furthermore, their analysis results in
a limit of $0.13 \times 10^6 M_{\odot}$ for the extended dark mass within $0.1$ pc
at $99.7\%$ confidence. A first robust estimate of this limit of the extended mass 
contribution was derived by \citet{Mouawad2005}.

In conclusion there is enough circumstantial evidence for the existence of a
supermassive black hole at the position of Sgr~A*. However, there have been no
attempts to estimate the mass at scales of the order of a couple of gravitational
radii. In the following we will outline a method using a relativistic model, 
which makes it possible to constrain the enclosed mass to within $25 r_g$.
If the method proves to be reliable, it is an appropriate way to test general
relativity in the presence of the strong gravitational field of a supermassive
black hole.

\begin{figure}
\centering
\includegraphics[totalheight=0.25\textheight]{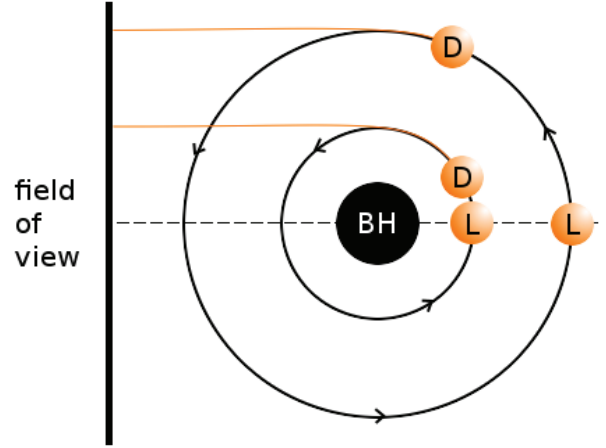}
\caption[something]{Illustration of the origin of the double-peak structure in
the total flux. The blobs marked with an 'L' are magnified by gravitational
lensing, while they are behind the black hole from the observer's point of view.
That is, they are positioned on the focal line, along the orbital section $Q$
outlined by the dashed line. 
The blobs marked with a 'D' are Doppler-boosted, because they are moving
'directly towards' (in terms of geodesics) the observer, as indicated by the
orange lines representing the geodesics from the source to the observer.
The 'field of view' is equivalent to the 'detector plane' in which the object is imaged.
\label{peaks}}
\end{figure}

\section{Description of the simulations}
\label{sec:simulations}

In this section we first give a brief summary of the simulations and then
justify and describe the methodology we use to 
model the brightest observed X-ray flares.
After summarizing essential facts on the radiation mechanism and 
characteristic features of the emission in different wave-bands, we
describe the details of the ray-tracing code that is used and
then discuss the setup of the model, 
which differs from the usual hotspot model.

\subsection{Summary of the method}
\label{subsec:summarymethode}
We fit the shapes of the brightest flares leaving the time-scale
in periods measured in gravitational units of $t_g=GM/c^3$, 
hence, leaving the mass as a free parameter.
Then the flare time-scale in periods is matched with the actual 
time length of the observed flare.
This allows us to derive black hole mass distributions from the simulations. 
The median values and median errors of these non-Gaussian distributions 
then correspond to the expected black hole mass and its uncertainty.
We apply the method to SgrA* and to the Seyfert~I galaxy RE~J1034+396.

\subsection{Flares}
\label{subsec:flares}

The discussion of the radiation mechanism shows that
the intrinsic flux density of the source component can be assumed to be
rather constant  during the rapid and strong modulations by relativistic 
effects like Doppler-boosting (beaming) or lensing.
Based on discussions in 
\cite{Neilsen2015, Eckart2012, Yuan2004, Baganoff2001, Markoff2001}
we conclude that the intrinsic flux density evolution of 
X-ray luminous components is dominated by the time-scales of the relevant 
sub-mm peaking synchrotron components.
This then implies that rapid flux density variations in the X-ray domain are
more likely a result of relativistic effects like boosting and lensing 
rather than synchrotron cooling due to high energetic electrons.
This is supported by the fact that the times scales for the boosting and lensing events
are very short (in the few minutes to 10 minute range - depending on the geometry)
compared to those on which the
underlying emission is varying. Also (again depending on the geometry) the amplification
factors of these relativistic effects can easily be several 10.
In addition, the flare lengths indicate that the flares themselves originate 
close to the black hole \citep{Baganoff2001,Baganoff2003}
making these relativistic modulations also frequent and relevant.
This scenario justifies that we model the flare profiles with an orbiting spot model.
This model is a generalized surrogate-model to characterize the behavior of (in this 
case) X-ray emitting matter in the gravitational field of a super massive black hole.
Of course, we only consider the variable non-thermal part of the SgrA* emission.
This source is embedded in an extended non-variable Bremsstrahlung component
\citep{Baganoff2001,Baganoff2003} and during times the source is not flaring,
the non-thermal flux drops well below the Bremsstrahlung flux level.
This constant component is therefore removed before we fit the flares.
Restricting the modelling to the brightest flares, therefore essentially 
completely avoids the risk of overlapping flare events, such that we truly model 
only one flare at a time.

\subsection{Ray-tracing code}
\begin{figure*}
\centering
\includegraphics[width=1\textwidth]{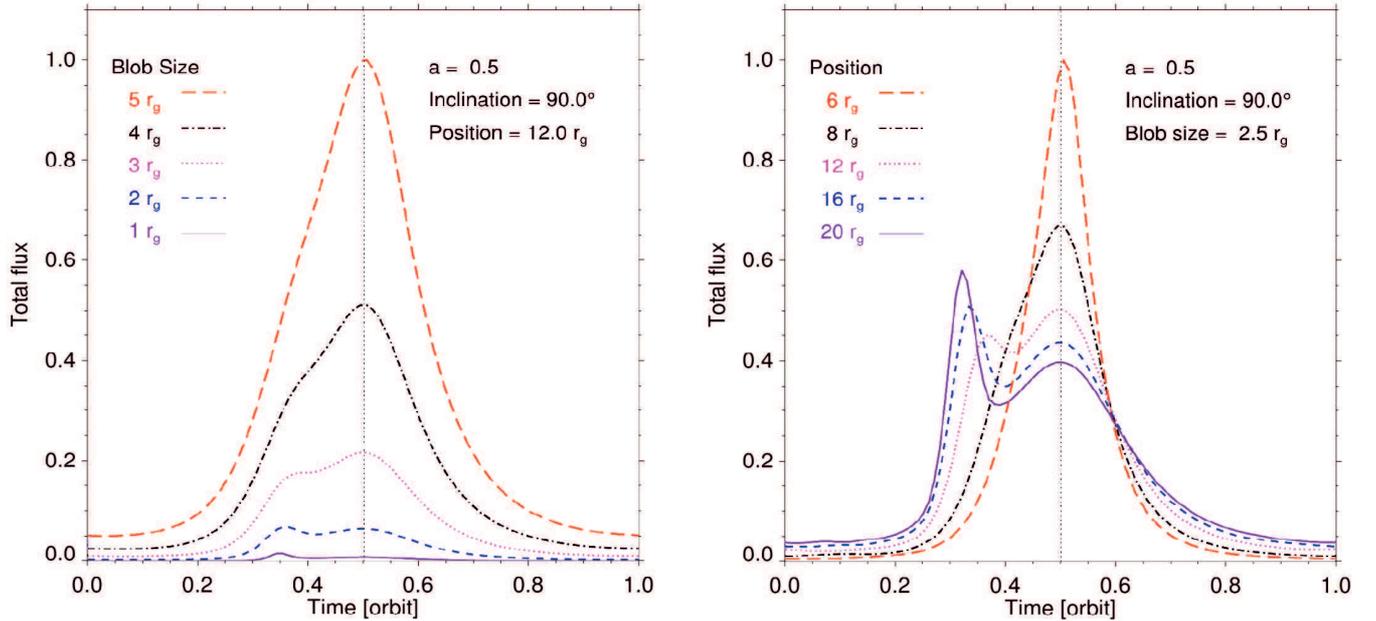}
\caption[something]{{\bf a)} Left: Illustrates the influence of the blob's size on
the shape of the light curve. The blobs of different sizes (5 $r_g$ red long
dashed, 4 $r_g$ black dash-dotted, 3 $r_g$ magenta dotted, 2 $r_g$ blue short
dashed and 1 $r_g$ solid purple line) are orbiting at a radial position of
12~$r_g$ around a black hole with spin 0.5, the viewing angle is 90~$^\circ$
(edge on).  The light curves are normalized
to the maximum of the peak value of the light curve for the blob with the size 5
$r_g$ and shifted such that the Doppler-peak is at the center.
{\bf b)} Right: Illustrates the influence of the blob's position on
the shape of the light curve. The blobs are orbiting at different positions (6 $r_g$ red long
dashed, 8 $r_g$ black dash-dotted, 12 $r_g$ magenta dotted, 16 $r_g$ blue short
dashed and 20 $r_g$ solid purple line) and have a size of
2.5~$r_g$ around a black hole with spin 0.5, the viewing angle is 90~$^\circ$
(edge on).  The light curves are normalized
to the maximum of the peak value of the light curve for the blob with the size 5
$r_g$ and shifted such that the Doppler-peak is at the center.}
\label{fig:fig2}
\end{figure*}

We introduce a numerical code based on the Kerr-metric library \texttt{sim5lib} 
written in the programming language $C$.
The code is also capable to describe polarization transport. However, unfortunately there is
no polarization data available for the flares we investigate here. A future X-ray
polarimeter might record an abundance of information that could help constrain
the black hole's parameters.
In an upcoming paper these polarization simulations will be discussed further
and applied to the near-infrared. However, here we restrict ourselves to the
total flux of Sgr~A* at X-ray wavelengths.
The library \texttt{sim5lib} uses the mechanism outlined by \citet{Dexter2009}, so
instead of using a set of pre-calculated transfer-functions on a grid of
parameters and then interpolating, as is the case, i.e., for the
Karas-Yaqoob~(KY) code \citep{Dovciak2004,Meyer2006a,Meyer2006,Zamaninasab2010,
Zamaninasab2011}, we use \texttt{sim5lib} to explicitly compute the null
geodesics and then track the emission along them using a parallel transport.
The null geodesics are
obtained by numerically solving the equation governing photon motion in Kerr
space-time. Starting at a maximum domain radius $R_{max}$, which is far enough
away from the black hole so that any general relativistic effects from there to
the observer are negligible, and going in the direction of the BH, we look for
those geodesics which reach an emitting source and discard the rest. Once these
relevant geodesics are found, the emission can be traced from the source back to
the observer in Boyer-Lindquist coordinates. Although this method demands much
more computing power than using pre-calculated transfer functions, it allows us
to control all parameters of the geodesics while the general relativistic
effects on the rays in the presence of a strong gravitational field are included
intrinsically. This comprises changes of the emission angle, rotation of the
polarization angle and gravitational lensing.
During the passage of the hotspot behind the black hole, as viewed by the observer,
the high gravity of the black hole leads to bending of the geodesics. This
gravitational lensing effect is at its strongest on the focal line.
This is demonstrated in Fig.~\ref{peaks}. 
The fraction of the orbit between the boosting and lensing points varies 
with the radius of the orbit, owing to the
stronger bending of the geodesics close to the black hole. 
In Fig.~\ref{peaks} this fraction is highlighted by a dashed line parallel to the corresponding orbit.
The comparable figure (Fig.12) in \cite{Eckart2017}
also shows the corresponding light curve with the boosting and lensing flux peaks labeled.
For comparison see also the light curves, particularly in Fig.~\ref{fig:fig2}(b)

\def\arraystretch{1.3}
\begin{table}
\caption[Table of the parameters of the simulations]{Basic parameters of the simulations.\label{parameters}}
\begin{center}
\begin{tabular}{l l l}
    \hline
    \hline
      Parameter & Values & Description \\ \hline
\def\arraystretch{1.0}
      $a$ & 0.5 & BH spin \\
      $M$ & 1 & BH mass \\% \hline
      $i$ & 5$^{\circ}$, 10$^{\circ}$, ... , 90$^{\circ}$ & inclination \\
      $D_{0}$ & 0.5 $r_{g}$, 1 $r_{g}$,  ... , 5.0 $r_g$ & size of the blob \\
      $R_{0}$  & 6 $r_g$, 8 $r_g$, ... , 24 $r_g$ & blob's radial position \\
      $\phi_{0}$  & 90 & starting azimuth angle \\
      $n_{e}$ & 1 & electron number density \\
      $R_{max}$ & 40 $r_{g}$  & max. domain radius \\
      Nx & 150 & resolution in pixels \\
      frames & 100 & time frames per orbit \\
      step & 0.1 $r_g$ & step along geodesics \\
    \hline
\end{tabular}
\end{center}
\end{table}

Additionally, the special relativistic Doppler-boosting on the intensity of the
radiation which is caused by aberration \citep{Einstein1905} is taken into
account as well. While the source moves away from the
observer, the emission is reduced, whereas it is magnified for an
approaching source.
Finally, the strong gravitational field of the black hole
presents a potential well, which exerts another redshift on the radiation. We
take these two effects into account by introducing another factor, the g-factor
\citep{Dovciak2004}:

\begin{equation}
g = \frac{\nu_0}{\nu_e} = \frac{p_{0t}}{p_iU^i} = \frac{E_\infty}{p_iU^i},
\end{equation}

where $\nu_0$ is the frequency of the observed photons, $\nu_e$ the frequency of
the emitted photons, $p_{0t}$ the time component of the four-momentum of the
photon as measured by an observer at infinity, and $U^{i}$ the four-velocity of
the hotspot.

\subsection{Setup of the models}

Usually the hotspot scenario is employed to model a localized brightness
excess within an accretion flow. 
A hotspot could arise through magnetic turbulence in a magneto-hydrodynamic
accretion flow \citep{Balbus1991,Armitage2003}, vortices and flux tubes
\citep{Abramowicz1992}, magnetic flares \citep{Poutanen1999,Zycki2002}, 
interactions of stars with an accretion disk \citep{Dai2010}, or magnetic 
reconnection \citep{Yuan2009}.
According to \citet{Eckart2012} it is most likely that the X-ray flares of Sgr~A* 
\citep[see also][]{Eckart2002, Baganoff2001}.
are caused by a synchrotron self-Compton process.
However, in this paper we only model the light curves resulting from the hotspot (or 'blob') motion
and not the physical mechanism that leads to the emission.
The idea is that instead of an accretion disk, there are several clumps of matter
of different sizes on different orbits with changing radii and viewing angles. 
These blobs of matter could have a similar origin as in the scenario of 
\citet{Jalali2014}, only on a much smaller scale. A cloud could be compressed
by the black hole's strong gravity and then quickly disrupted. However, in the
following only enhanced blob luminosity over half an orbit is needed. 
This is sufficient to distinguish between different relativistic effects. 
A cloud that is below the detection limit, could be magnified by the relativistic
effects strongly enough such that it results in a flare, when viewed by the
observer.
The variability of Sgr~A* would in this sense result from a physical one-state
statistical process, as was argued by \citet{Witzel2014}.

\subsubsection{The modelling parameters}
The important parameters that are varied in the simulations are the blob's size $D_0$, the
radial position of its orbit $R_0$ and the blob's orbital inclination $i$ with
respect to the observer.
An overview of all the parameters of the model can be found in
Table~\ref{parameters}.
Most of the other parameters are of lesser importance 
and - for the purpose of the fit done in periods and gravitational units $t_g$ - can be set to 
constant values as described below.
This is the minimum set of parameters required to describe the motion of
an emitting source in the gravitational field of a super massive black hole.

To model the hotspot we use a three dimensional Gaussian which emits uniformly
in all directions, and is assumed to be in Keplerian motion in a stable orbit
around a black hole.

\begin{equation}
K = n_e \, \mathrm{exp}\left[-\frac{1}{2}\frac{x^ix_i+\left(x^iU_i\right)^2}{D_0^2}\right],
\end{equation}

where $n_e$ is the number density of electrons, $x_i$ the vector difference
between the position of the photon and the center of the blob, $U_i$ the
Keplerian four-velocity of the plasma and $D_0$ is a measure of the size of the
blob, given in gravitational units. In our simulations we can safely assume
$n_{e}$ to be 1, because in our model the number density is only a scaling
factor for the luminosity.

Nothing is assumed about the mass of this black hole: all computations are
undertaken in gravitational units. This will allow us to infer constraints on
the black holes' mass after the fitting process.
Apart from determining the innermost stable
circular orbit (ISCO) of the model, the spin of the black hole $a$ does not
exert any noticeable influence on the shape of the light curves. Here, we only
consider a black hole with prograde spin, that is $a \in [0,1)$, and in the
    orbiting hotspot model the spin is a variable of the period of the orbit
 ($G=c=1$):
\begin{equation}
P_g = 2\pi [R^{3/2}+a]~~~,
\end{equation}
where $P_g$ is in geometrical units, and $R$ is the radius of the orbit, measured in units of the gravitational radius.
Alternatively, we can write \citep{Dovciak2004}:
\begin{equation}
P_T = 310 (R^{3/2}+a) M_7 = 49.3 P_g \times M_7~~~,
\label{equ:PT}
\end{equation}
with $P_T$ in seconds and the mass $M$ in units of $10^7 M_{\odot}$.
Thus we choose $a=0.5$ and in doing so accept an uncertainty of $\pm \pi$
of the period. The lowest radius which is taken into account using 
$R=6r_g$, 
with $M=0.4\times M_7$, $a=0.5$ and $R=6r_g$, leads to
a period $P_T \sim 31 min$. From equation~\ref{equ:PT} we can see that 
the $\Delta a= 0.5$ uncertainty of the spin results in a
relative uncertainty of $\pm 3\%$ of the period.

In our simulations we consider various inclinations of the hotspot orbit (see Tab.~\ref{parameters} for the range)
with respect to the spin axis of the black hole.
The viewing angle is measured from the rotational axis of the black hole, so when
we assume that the spin axis remains the same while we vary the viewing angle (i.e. the inclination  of the orbit),
another uncertainty is introduced owing to the precession due to the Lense-Thirring effect \citep{Lense1918}.
According to \citet{Merritt2010}, the precession time-scale is given as
\begin{equation}
    T = \frac{P}{4 a} \bigg[\frac{c^2 a_{sma} (1-e^2)}{GM}\bigg]^{3/2},
\end{equation}
where $P$ is the period in physical units, $a_{sma}$ the semi-major axis and $e$ the eccentricity of the orbit.
Here, we are assuming a circular orbit, thus we have
\begin{equation}
    T = \frac{P}{4 a} \bigg[\frac{c^2 R }{GM}\bigg]^{3/2},
\end{equation}
or, with $G=c=1$ and $a=0.5$,
\begin{equation}
    T = \frac{P}{2} R^{3/2}.
\end{equation}
For the case of the smallest radius taken into account $R=6 \times r_g$ this timescale is
more than seven times that of the period of the orbit. For all radii which are bigger,
it is even more than that. Thus, precession effects on the inclination can be
neglected in our considerations, especially considering that the step width between
inclinations is $5^{\circ}$.

\section{The Simulations}

%Therefore, each simulated flare is a 
%combination of two
%peaks which blend into each other \citep[cf. also][]{Dexter2009}.
%In general, the biggest magnification is achieved at high inclinations. This is
%because gravitational lensing and the Doppler blue-shift have a bigger influence at
%high inclinations. When the disk is completely face-on, the plasma blob never
%moves towards or away from the observer, and so there is no Doppler-shift and
%no modulation of the intensity. Similarly, in order for the black hole to act
%as a gravitational lens, the hotspot must be on the focal line. Hence, the blob
%needs to be at least a little bit 'behind the black hole' with respect to the
%observer, which is impossible in a face-on scenario.
%In this regard, magnification of the emission by gravitational lensing is more
%sensitive to the viewing angle of the observer than magnification by
%Doppler-boosting, due to the sensitivity of the lensing near the focal line.
The relativistic effects which are predominantly responsible for the
magnification, gravitational lensing and the Doppler-boost, occur within 
within {\it less than} a fraction ($Q$) of an orbit.
This matches well with the life time estimates of the orbiting spots
(see Fig.~\ref{peaks}).
Theoretically, an accretion disk spot is assumed to rarely last for
much longer than about one orbit 
\citep{Schnittman2005,Schnittman2006, Meyer2006a, AdamsWatkins1995}.
In fact, a third of an orbital time-scale is sometimes indicated
\citep[e.g. see discussion in section 5.1.1. by][]{Eckart2008,Schnittman2006}.
This time-scale is also well matched by our full width half
maximum flare lengths of about 0.3 obits in Fig.~\ref{peaks}.
Depending on the distance of the hotspot from the black hole this
then covers the observed overall flare time-scales very well
and is also compatible with the time-scale for the intrinsic 
variation of the emission as discussed previously.
Both constant 
\citep[e.g. Fig.1 in][]{Abramowicz1991}
and exponentially decaying 
\citep{Schnittman2006}
light curves are assumes in the
literature.
Assuming an exponential decay of the spot flux density 
over the characteristic time-scale $t_{lifetime}\sim 0.3 t_{orbit}$ 
then the drop will only be less than 50\% over the section $Q$ 
and only 25\% over the actual boosting and lensing phases.
This result can be applied to SgrA* since all 
theoretical magneto-hydrodynamic accretion models 
show a so called 'central mid-plane' which is comparable to a disk
\citep{Moscibrodzka2009, Moscibrodzka2013, Moscibrodzka2014}.
We consider the hotspot as the dominant part of a much 
fainter disk component that we do not model.
This also supports the choice of spots on circular orbits as a 
surrogate model for radiating matter close to the SMBH
as highly elliptical orbits due to infalling matter are 
probably strongly suppressed.
In a viscous environment with multiple gaseous clouds  
(as expected for 'central mid-planes' resulting 
from magneto-hydrodynamic accretion models),
clouds on crossing orbits can be excluded as their collisions
are highly dissipative.
As (semi-)stable trajectories in such an environment circular 
orbits are preferred.

Also, at least for SgrA* we do not claim repeated orbital periods mainly
for the reasons above and due to the lack of observational evidences.
However, the situation could be more complex as can be seen from the example
J1034-396 that we refer to towards the end of the article.
If the light curve of this source is interpreted using an orbiting 
spot model then multiple orbital periods and
longer spot life times could be involved.

\begin{figure*}
\centering
\includegraphics[width=\textwidth]{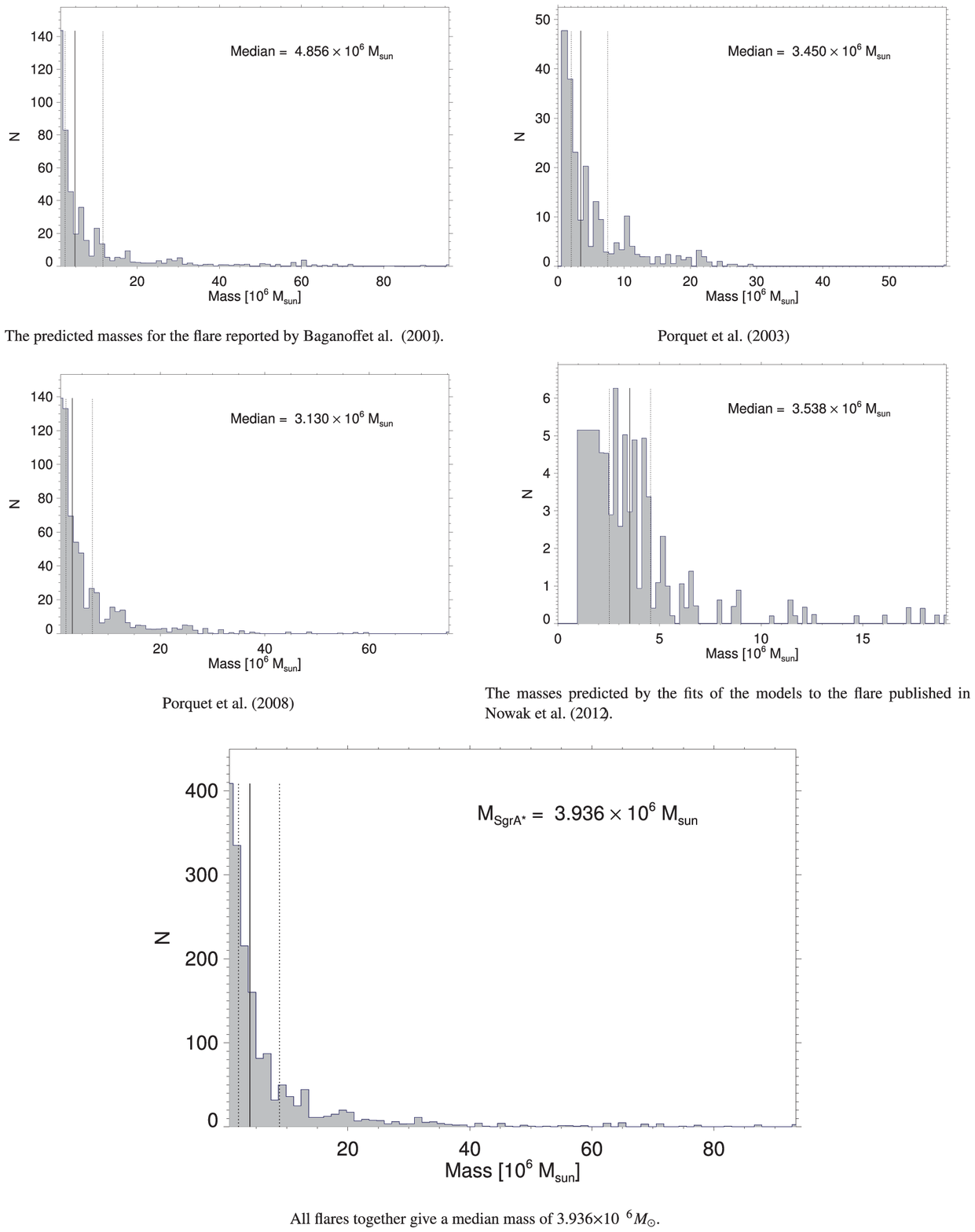}
\caption[something]{Weighted histograms of the predicted masses for all the
    models for the four flares taken into account.
 \label{masses}}
\label{Weight}
\end{figure*}

\begin{figure*}
\centering
\includegraphics[width=\textwidth]{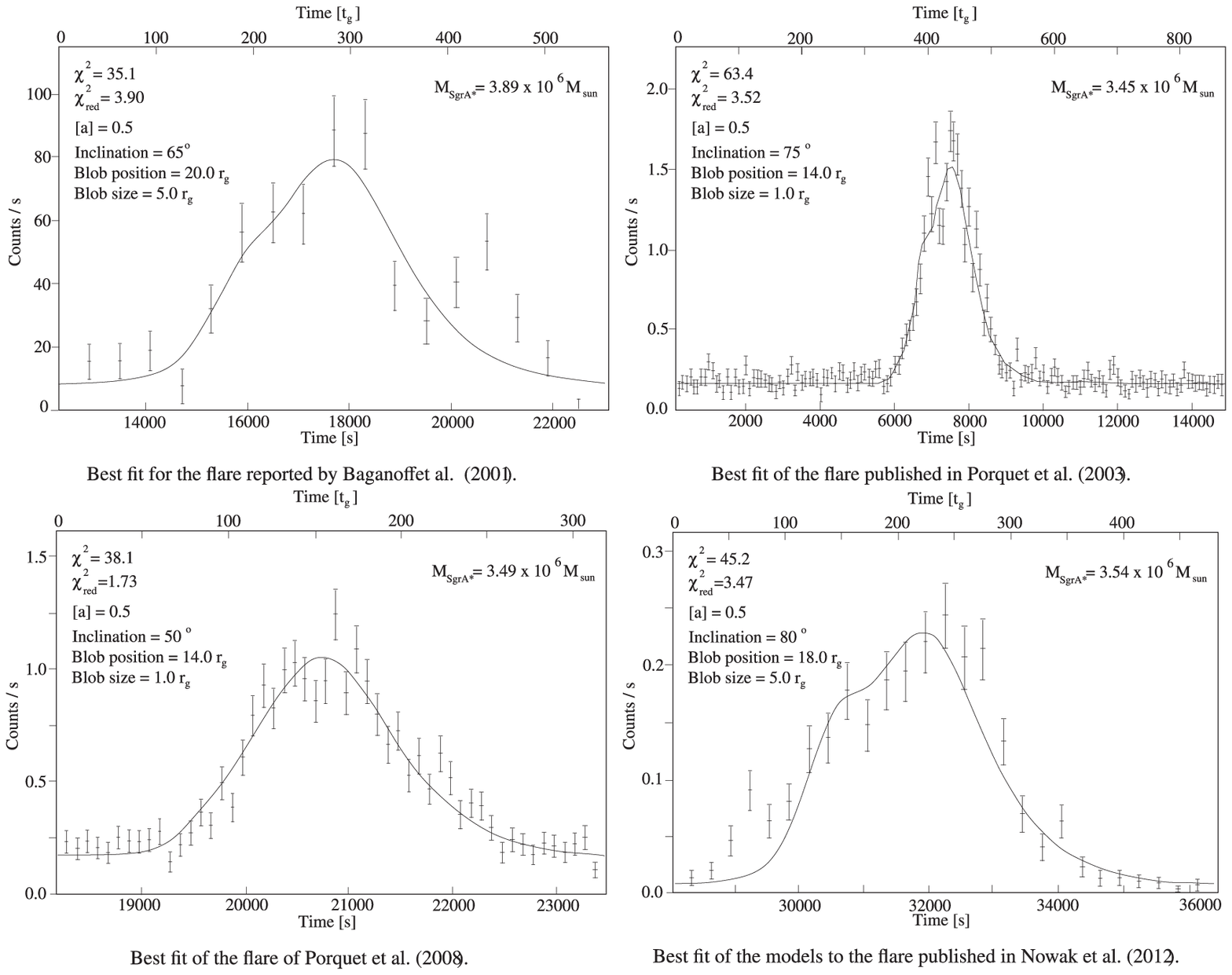}
\caption[something]{Best fits for the four analyzed flares in their respective median bin. \label{bestfits}}
\end{figure*}

\citet{Broderick2006} find that the size of the hotspot indeed does not have a dominant
influence on the light curves.
 However, as can be seen in Fig.~\ref{fig:fig2}ab, the
simulations with \texttt{sim5lib} show a clear dependency of the shape of the
flare on the size of the plasma blob, as well as on the distance of the blob to
the black hole.
In several previous ray-tracing simulations with a two-dimensional
hotspot \citep[e.g.][]{Meyer2006a,Meyer2006,Zamaninasab2010,Zamaninasab2011}
this relation was not apparent. This is because high inclinations are needed in 
order to see this effect, which is problematic with a two-dimensional source.
\citet{Mossoux2014} noted that the ratio between those two peaks is influenced
by the blob's size, which is confirmed by our simulations.
We interpret this to arise because of the specifically different natures of the two relevant
effects,  which favor different conditions for the highest possible relative
magnification. For a small emitting region (i.e. a blob
size of $\sim0.5~r_g$), the magnification due to lensing is much more
effective than the magnification due to the Doppler-effect. A sufficiently
large hotspot (i.e. a blob size of $\sim5~r_g$) however, will magnify only slightly
by virtue of the black
hole's gravitational lensing effect, in comparison to its much more efficient
magnification via the Doppler-effect.

Furthermore, the radius of the orbit has a major influence on the profile of the
light curves, as can be seen from Fig~\ref{fig:fig2}. In particular, the size of
the hotspot itself is not the only
factor which influences the ratio of the two peaks: the further a blob is
away from the black hole, the bigger the blob needs to be in order to result in
a dominant
magnification of the Doppler-effect. While on the other hand, a blob close to the
black hole requires a particularly small blob in order to yield a dominant
gravitational lensing magnification.

An additional effect that influences the light curves' shapes is dependent on the
radial position as well. For blobs on a close orbit around a black hole, the
geodesics are bent so much that the
Doppler-peak immediately follows the lensing-peak, whereas for larger orbits
the peaks can be as much as a quarter of an orbit apart. This effect and the
mechanism by which the two peaks originate are illustrated in
Fig.~\ref{peaks}.
This means that it is generally possible to get a sharp drop to almost the
quiescent state level, i.e. below the detection limit, between the two peaks by
using a very small hotspot orbiting at a sufficient distance from the black hole.

Structures of the flares that are not due to flaring and boosting will have the
tendency to be interpreted as smaller source sizes and larger distances 
(see Fig.~\ref{fig:fig2}ab).
For small (10-20\%) variations this will have little effect on the mass estimate
as larger distances correspond to lower velocities. However, the overall
quality of the fit may be affected.

\def\arraystretch{1.3}
\begin{table*}
\begin{center}
\begin{tabular}{l l l l l l l}
    \hline
    \hline
    Flare & $M [10^6M_{\odot}]$ & $i [^\circ]$ & $R_{0}[r_g]$ & $D_{0}[r_g]$   & Amp$_{max}$\\ \hline
       \cite{Baganoff2001} & $4.86^{+6.80} _{-2.41}$ & $46.47$ & $15.69$   & $2.85$ & 25\\% \hline
       \cite{Porquet2003}  & $3.45^{+4.07} _{-1.43}$  & $55.40$ & $12.43$  & $2.59$ & 39\\
        \cite{Porquet2008} & $3.13^{+3.81} _{-1.24}$ & $49.16$ & $14.24$    & $2.84$ & 32\\
          \cite{Nowak2012} & $3.54^{+1.02} _{-1.01}$ & $69.52$ & $17.90$    & $3.60$ & 51\\
             Median  & $3.49 \pm 0.20$ & ~ & ~ & ~ \\
             All flare fit  & $3.94^{+4.85}_{-1.86}$ & ~ & ~ & ~ \\
       \cite{Mossoux2014} (Epic)& $3.18^{+5.56} _{-2.57}$ & $60.68$ & $14.14$  & $3.06$ & 49\\
    \hline
\end{tabular}
\end{center}
\caption{The median values of the mass model parameters
   $M$, $i$, $R_{0}$, and $D_{0}$ as well as the corresponding maximum amplification factor Amp$_{max}$.
\label{mean-parameters}}
\end{table*}

\section{The fitting routine}
In this section the fitting process we use to infer an estimate on the mass of
Sgr~A* is described in detail.
Here, we concentrate on the sections of the light curves that
contain the boosting and lensing model information. We also
insure the comparability of the fitting results obtained for flares
with different sampling and signal to noise ratio.
First the simulated light curves are normalized such that the maximum flux of
the light curve $f_{sim,max}$ is the same as the one of the observed light curve
$f_{obs,max}$. Then for each light curve the ratio between $n_{f}$ (the number of
data points of the light curve which belong to the flaring period - here defined
as data points with a flux above 30\% of the maximum) and $n_{q}$ (the number of
data points of the quiescent state) is considered.
As shown in Fig.~\ref{fig:fig2} all major features of the flares that depend on the relevant 
model parameters are contained in the upper two thirds of the flares.
Every simulated light curves' ratio is then compared to an observed one and
quiescent state data points are removed from or added to the simulated data
until the ratio is comparable. 
This is possible and necessary since the simulated light curves always have a better 
time resolution.
Since the density of data points per time-equivalent interval remains the same 
this method is fully equivalent to 're-binning'.
It has, however, the advantage of being insensitive against binning phase shifts between 
the simulated and observed data.
By controlling the ratio $n_f$/$n_q$ between on-flare and off-flare (i.e. on baseline) data points 
we insure that the $\chi^2$-values of the mass estimates from different flares and fits remain comparable.

To conduct a time efficient fit of the models to the data we introduce 
a time shift, a flux density scaling factor, and a flux density offset.
The best time shift, $t_{shift}$, needs to be found as the light
curves do not necessarily have the peak at the same position. 
This fit is done in a sufficiently large window (plus minus one quarter of the flare length) 
as the expected separation of the two peaks in the double-peak flare structure structure of the light curves
can be at most a quarter of an orbit apart - as can be seen in Fig.~\ref{peaks}.
The simulated data is multiplied by a factor, $f_{scale}$,
because in general a best fit of the shapes of the light curves does not depend on the
initial normalization we inferred earlier. 
A constant, $c_{bg}$, is added to or subtracted from the flux of the simulated
data to account for residual offsets of the baseline with respect to which we investigate
the flare.
The $\chi^2$ between a simulated light curve and an observed light curve is
calculated on all data points of the flaring part as defined above:

\begin{equation}
\chi^2 = \sum_{t} \frac{\bigg(f_{obs}(t) - f_{scale}\times f_{sim}(a, i, R_0, D_0, t+t_{shift}) + c_{bg}\bigg)^2}{\Delta f_{obs}},
\end{equation}
With the reduced $\chi^2$
\begin{equation}
\chi^2_{red} = \frac{\chi^2}{n_q+n_f-3},
\end{equation}
where the number of parameters is three, because the only parameters varied
for this fit are $t_{shift}$, $f_{scale}$ and $c_{bg}$. This defines the best
fit of a particular model to a particular observed flare.
Otherwise, we conduct a full grid search covering 
In the model parameters
$i$, $R_{0}$, and $D_{0}$ with step sizes giver in Tab.~\ref{parameters}.

However, to compare the fits of the simulated flares to a particular observed 
flare, only the data points of the flaring period are taken into account, in
order to avoid artificially improving the fit by fitting quiescent state data
points. Thus we define a second $\chi^2$:
\begin{equation}
\chi^2_{flare} = \sum_{t \in T_{flare}} \frac{\bigg(f_{obs}(t) - f_{scale}\times f_{sim}(a, i, R_0, D_0, t+t_{shift}) + c_{bg}\bigg)^2}{\Delta f_{obs}},
\end{equation}
and the accordingly reduced $\chi^2$:
\begin{equation}
\chi^2_{flare,red} = \frac{\chi^2}{n_f-3}~~~.
\end{equation}
In this case the number of parameters is three as well, because we vary the
inclination $i$ or the orbit to the observer, the radius $R_0$ of the orbit and the size $D_0$ of the hotspot.
In summary: $\chi^2$ is the raw non-reduced value,
$\chi^2_{red}$ describes the quality of the fit to the data. 
Hence, these values are also shown in the corresponding panels in which we show the fitted data.
The value $\chi^2_{flare,red}$ describes the weight one can attribute to the physical model 
parameters that lead to the fit. This value is also used to weight the resulting mass estimate.

Note that our main objective here is not to find a model that is an exact fit
for any of the observed flares, but rather to investigate whether or not
constraints can be put on the mass of the black hole.
Each fit gives an estimate for the black holes' mass, which is calculated from
the lengths of the simulated and observed light curves. This is possible because
the duration of an observed light curve is naturally known in seconds, whereas
the duration of simulated flares is known in gravitational units. 
Both are linked to each other via the black hole mass.
Due to the fitting process the theoretical light curves are lined up with the observed ones
and the conversion factor between gravitational time units of a particular model
and the timescales of a particular observation  in seconds can be calculated.
This factor is only dependent on the mass of the black hole, which is then given
by:
\begin{equation}
M_{SgrA*} = \frac{c^3}{G M_{\odot}} \times \frac{T_{obs}}{T_{sim}},
\label{equ:mass}
\end{equation}
in units of $M_{\odot}$, where $T_{obs}$ and $T_{sim}$ are the durations of the
observed and simulated light curves in seconds and gravitational units $t_g$,
respectively. By gauging the intrinsic clock of the black hole in its
gravitational units to the clocks of the observations in seconds, the mass of
the black hole can be estimated.

 Usually, for the light curve calculations the black hole mass is inferred. 
Combined with the duration
 of an observed flare is sufficient to estimate how close the clump of matter,
 which is responsible for the flare, is to the black hole in terms of
 gravitational units.
 However, the mass of Sgr~A* is left as a free parameter, and radii between
 $6r_g$ and $24r_g$ are tested. 
The choice of this range does not pre-prompt the probed masses as the
flare lengths measured in periods or gravitational units $t_g$ need to be calibrated 
first with the flare length measured in the observer's frame before getting 
a mass.
In the case of SgrA* this range is consistent with the radio size measurements. 
According to \citet{Reid2004}, Sgr~A* has an
 intrinsic size of 1 AU. This corresponds to $\sim 25 r_g$ in case of a black
 hole with a $4 \times 10^6 M_{\sun}$ mass. If Sgr~A* has a bigger mass, our
 range of parameters is already outside of its known size. However, if its mass
 is considerably smaller, Sgr~A* could in principle be bigger than $25 r_g$.
Also, there is no tendency for a degeneracy in terms of orbital time-scales,
say between 
a blob with a full orbit of a blob at a radius of 13$r_g$
and
half an orbit of a blob at radius of $2^{2/3}$$\times$13$r_g$=20$r_g$,
since the relativistic flare modulation is mainly done along the 
orbital section in which the boosting and lensing is taking place
(see Fig.~\ref{peaks}).

\section{Results}
In this section we present the results of the fitting process for the four
analyzed flares and analyze the resulting masses. Then we use our method on
another flare of Sgr~A* published by \citet{Mossoux2014} and on a 
different source to test whether or not it is a reliable method.

\subsection{Mass estimates}
Each fit of a particular simulated light curve to a particular observed flare
results in an estimate for the black hole mass as outlined above.
Note that we are not trying to fit the light curve perfectly but that we are only 
interested in the predicted masses: 
we employ the use of a toy model of an orbiting blob or hotspot without taking into
account the radiation mechanism. 
Only masses resulting from a fit that
 has a $\chi_{flare,red}^2<5$ are taken into account.

First we consider each flare individually as a single event and do not assume
that they are connected to each other in any way.
 In Fig.~\ref{masses} the resulting masses predicted
 for the four flares by all the models are displayed on the $x$-axis, while the
 $y$-axis represents the number of model results which predict the particular mass.
Each result has been weighted by $1 / \chi_{flare,red}^2$, i.e. the $y$-axis label number $N$ 
represents the weighted number of mass estimate per mass bin.
These figures give an indication which mass is
 predicted for most models that are a good fit. A peak in this diagram occurs
 when most models, which predict a mass in the particular mass bin, have a small
 $\chi_{flare,red}^2$. 
Since the resulting distributions are not normal distributions, we use the medians 
(indicated by the vertical solid lines)
as a measure for the mass. The asymmetric (i.e. calculated separately for each side) 
median deviations from the medians are
 calculated as the median of the deviations to left or right of the medians.
(indicated by the vertical dotted lines).
 The estimates for the masses, and the median values for the models which
 contribute to the biggest peak in these
 diagrams are displayed in Table~\ref{mean-parameters}.
These models contain the best fits shown in Fig.~\ref{bestfits}.
In this table the median variations of the
inclination $i$ are $20^\circ$, of the orbit radius $R_{0}$ they are $2~r_g$,
and for the source diameter $D_{0}$ we find a variation of $1.5~r_g$.
The maximum amplification factor Amp$_{max}$ may vary between 40\% and 60\%.
In the modelling, small offsets due to the none boosted emission 
of the blob will be compensated for via the constant offset $c_{bg}$. 
In reality, these small offsets will be irrelevant as the live time of 
the blob is only about one third of the orbital time-scale.
As a median  mass value of the medians of the 4 individual flares we obtain
$M_{SgrA*, median} = (3.49\pm0.20) \times 10^6 M_{\odot}$.
 The resulting masses differ from flare to flare. 
However, the flares should arise
from the vicinity of the same black hole and its mass
should not change in such a drastic way. 
Hence, at the bottom of  Fig.~\ref{masses} 
we plot another weighted histogram
in which we take into account the masses which result from the fits of all flares.
As a median mass we find here 
$M_{SgrA*, all} = 3.94^{+4.85}_{-1.86} \times 10^6 M_{\odot}$.
For each histogram we find the best fit that lies in the bin of the median of the
distribution. 
They best fits are plotted in Fig.~\ref{bestfits}.
and the corresponding mass histograms are plotted in Fig.~\ref{Weight}.
This must be compared to the current stellar orbit based  mass estimate 
of $M_{SgrA*, stars} = 4.02 \pm 0.17 \times 10^6 M_{\odot}$
by, e.g., \citet{Boehle2016}.
The median mass estimate $M_{SgrA*, median}$ agree to within three times the uncertainties,
and the $M_{SgrA*, stars}$ value is well included in the range provided by the $M_{SgrA*, all}$ estimate.

\subsection{Application to a faint SgrA* flare}
\begin{figure}
\centering
\includegraphics[width=0.48\textwidth]{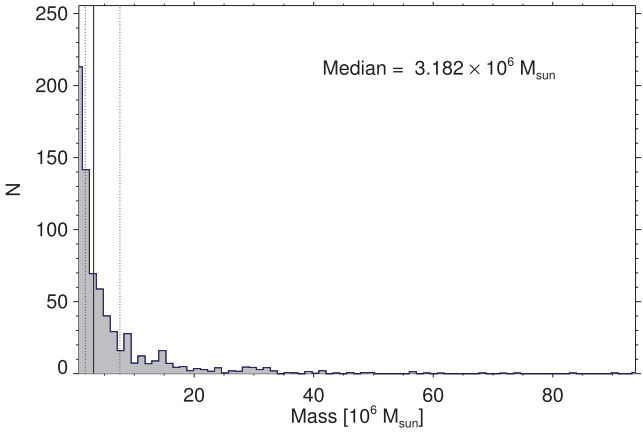}
\caption[something]{Predicted masses for the flare published by
  \citet{Mossoux2014}, detected with XMM/Newton/Epic instrument.
  \label{mossoux_hist}}
\end{figure}

\begin{figure}
\centering
\includegraphics[totalheight=0.25\textheight]{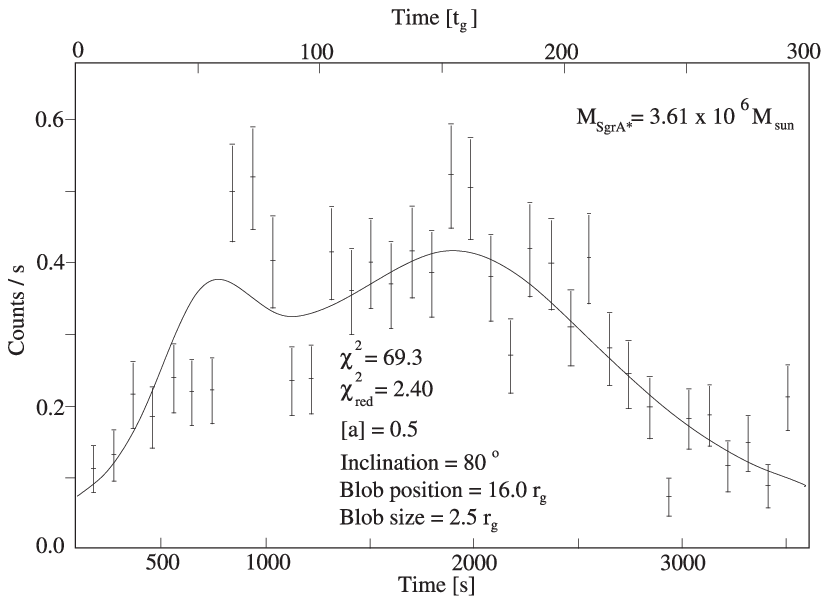}
\caption[something]{Best fit for the flare published in
\citet{Mossoux2014} \label{mossoux_bf}}
\end{figure}

Faint flares are statistically more frequent than bright flares. Hence, there
may be overlaps between different flare events and the method cannot be applied 
straightforwardly.
However, the flare shape may be indicative for an event 
dominated by the effects of relativistic motion of the luminous matter orbiting 
the black hole.
Therefore, we also investigated the flare published by \citet{Mossoux2014}, which is not as
bright as the 4 flares we used above, however, it shows two distinguishable peaks
as well. We only take the data from the measurements that were taken with the
XMM/Newton/Epic instrument, as it shows the highest count-rate. 
Our mass analysis is shown in Fig.~\ref{mossoux_hist} and results in a median of
$3.182 \times 10^6 M_{\odot}$. The best fit of the median bin gives a mass estimate
of $3.61 \times 10^6 M_{\odot}$, as can be seen in Fig.~\ref{mossoux_bf}.

\begin{figure*}
  \centering
    \includegraphics[width=\textwidth]{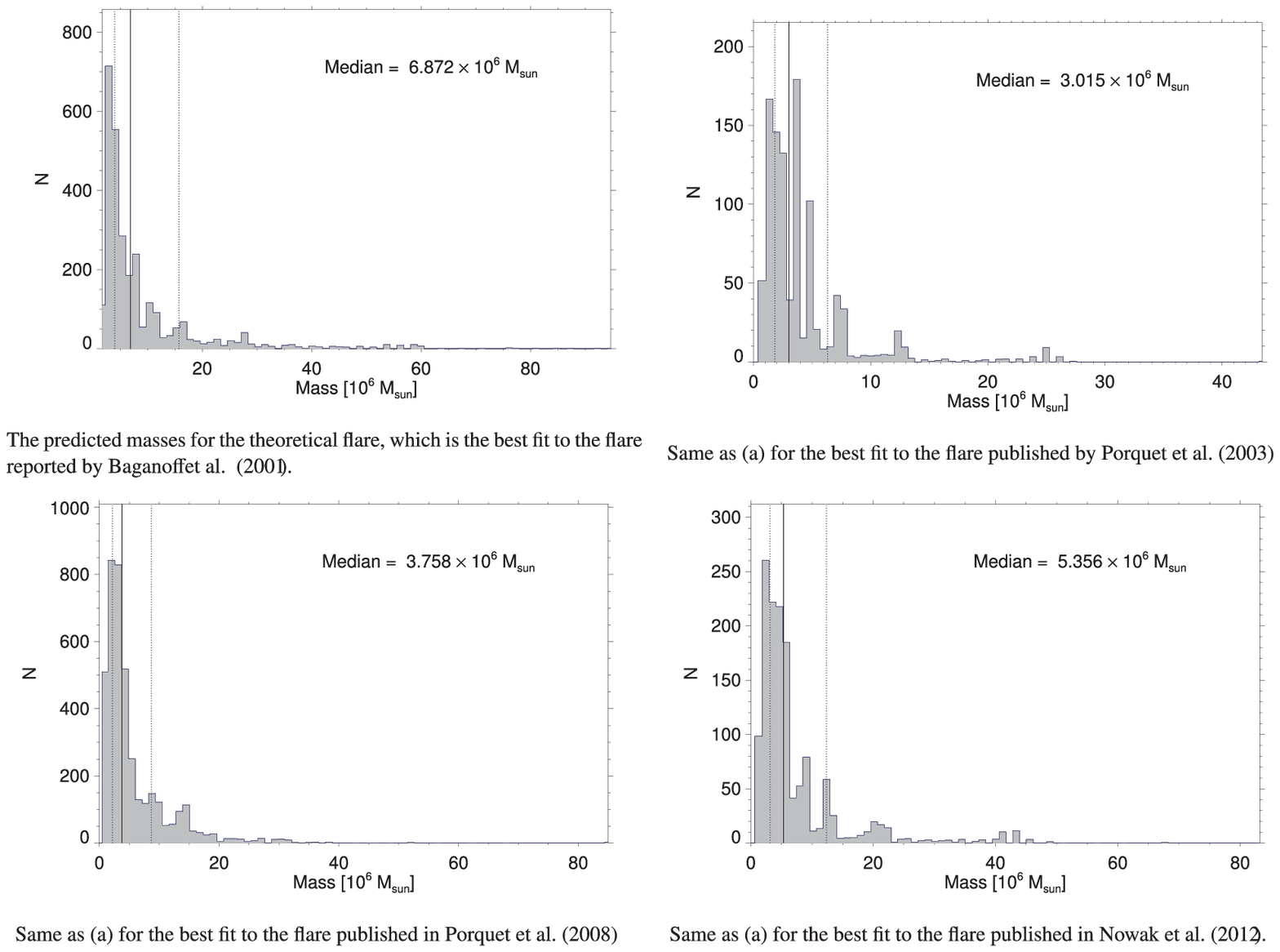}
  \caption[]{ \label{fitself}}
\end{figure*}

\subsection{Robustness of the method}

The question remains, whether the median of the distribution of obtained
is actually a good
measure for the mass of the black hole or not. 
Therefore, we tested the method against artificial light curves and 
applied it to data of a different, extragalactic supermassive black hole.

\subsubsection{Artificial flares}

To investigate how accurate the proposed method of constraining the black hole
mass is, we test the method against light curves that are produced by a 
black hole with a known mass. 
As we are probing models with the same routine we generate the test
models with - this can only be a test of whether the median of the 
mass distribution is appropriated to derive the model input mass again.
In this sense the test is only a consistency test of the input versus the
output of the model.
We used the best fits to the four flares, set the associated mass to
a known value of $4 \times 10^6 M_\odot$.
Here, it is only important is that a value was set. 
The results will scale with any mass value set differently.
This leaves us with four theoretical light curves
corresponding to a black hole of a known mass. Furthermore, each one of these
light curves are the best fit of one of the observed flares.
From these theoretical flares we extracted data on the same time support as the
observed flares and applied our method to these artificial flares.
The resulting mass diagrams are shown in Fig.~\ref{fitself}.
As a median  mass value of the medians of the 4 artificial flares we obtain
$M_{artif., median} = (4.56\pm0.80) \times 10^6 M_{\odot}$.
This agrees to within the uncertainties with the preset value of
$4 \times 10^6 M_\odot$.
Within 2 times the uncertainties it agrees with the 
$M_{SgrA*, median} = (3.49\pm0.20) \times 10^6 M_{\odot}$
result obtained for SgrA* with a measured mass of
$(4.02 \pm 0.17) \times 10^6 M_{\odot}$ by \citet{Boehle2016}, i.e., a value close to
the preset mass. 
We conclude that mass can be determined with  our new method with an accuracy 
between 5\% and 20\% (with a median value around 12\%), and that in addition to detector noise 
the step width of the basic simulation parameters listed in Tab.~\ref{parameters}
as well as the limited time support of the light curves
is a main source of the uncertainties.

\subsubsection{Applying the method to a different source}

We also test the method on another X-ray source which displays a
periodicity, RE~J1034+396, a Seyfert~I galaxy.
\citet{Gierlinski2008} and \citet{Middleton2009} have published observations of
this source, that show a QPO with a duration of about 1 hour at X-ray
frequencies. This QPO has further been confirmed by \citet{Vaughan2010}, who
states that the detection was not as significant.
\citet{Middleton2011} however, found no evidence of a periodic feature in the
power spectral density, but cannot exclude a QPO in two low-flux observations.
This prompts the conclusion, that the QPO might be a temporary feature.
\citet{Alston2014} analyzed several observations of this source and found that
this quasi period oscillation can be observed in five XMM-Newton observations.
The authors further state, that the QPO can only be found when the source is in
a low-flux state.
  Several authors have given estimates for the mass of this source
\citep{Gierlinski2008,Bian2010,Jin2012},
which are listed in Table~\ref{seyfert}.
The fact that this source appears to exhibit QPOs particularly during low-flux
phases, while it does not necessarily appear in all observations, suggests that
this feature might arise from a similar mechanism as the flares of Sgr~A*.

\citet{Czerny2010} find that the period of the QPO of this source appears to
increase with increasing flux. Following their timing analysis, the authors 
exclude a hotspot model on the basis, that only a face on hotspot could explain
the observations, which the authors do not find very probable.

\begin{figure}
\centering
\includegraphics[totalheight=0.25\textheight]{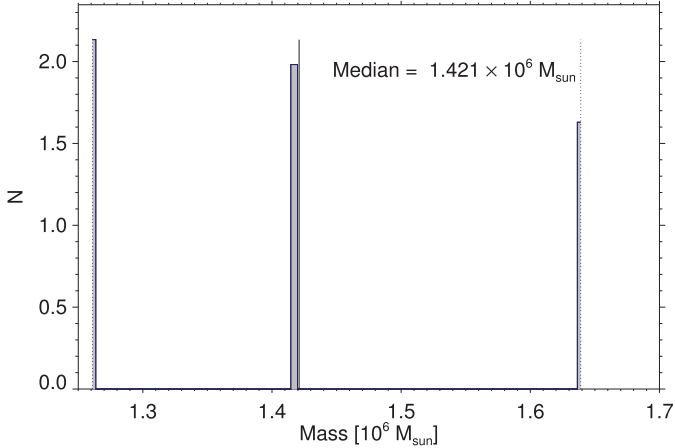}
\caption[something]{Weighted histogram for the QPO of J1034-396 as published by
\citet{Gierlinski2008} \label{hist_gierqpo}}
\end{figure}

\begin{figure}
\centering
\includegraphics[totalheight=0.25\textheight]{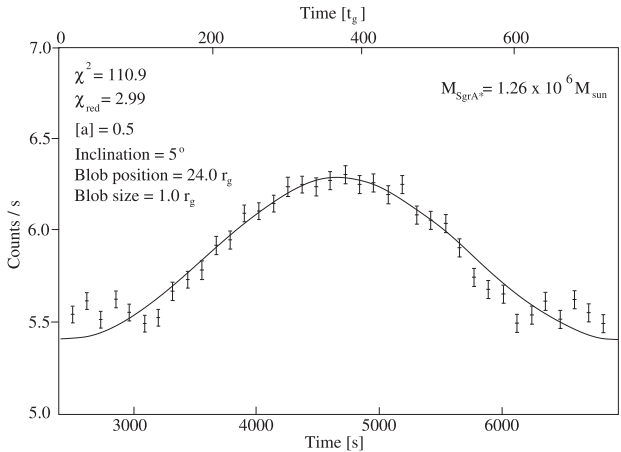}
\caption[something]{Best fit for the QPO of J1034-396 show for data of the 
folded light curve as published by
\citet{Gierlinski2008} \label{gierqpo}}
\end{figure}

We apply our hotspot model the same way as outlined for the X-ray flares of
Sgr~A*. The histogram of all resulting mass estimates is shown in
Fig.~\ref{hist_gierqpo} and the median mass is around
$M_{RE~J1034+396} = 1.421 \times 10^6 M_{\odot}$.
The best fit to the light curve is shown in Fig.~\ref{gierqpo}.
This mass is consistent with the estimates of other
authors, who have used different methods (cf. Table~\ref{seyfert}).
The fact that only models with a very low inclination result in an acceptable
fit, compares well with the observation that this particular source appears to 
only exhibit a QPO during low flux phases. As stated above, in our model the
biggest magnification is achieved at high inclinations, because then the special
as well as the general relativistic effects are most effective.
The QPO of RE~J1034+396 is actually only a modulation of $\sim 15 \%$, whereas
X-ray flares of Sgr~A* can reach fluxes that are magnified by a factor of $>100$,
when compared to the quiescent level. That is the reason why all flares of Sgr~A*
that were analyzed, compare better to models of high viewing angles, while the
QPO of RE~J1034+396 most likely results from a hotspot orbiting at a face on 
viewing angle with respect to the observer. Indeed, there are only three fits with
a $\chi^2_{flare,red} < 3$ in our analysis of this source, all of which are at a viewing angle
of $5^\circ$. This is consistent with the conclusions of \citet{Czerny2010}, even
though we have employed a completely different method.

We conclude that this new way of estimating the mass of black holes gives results
that are in agreement with those of independent methods. Not only is the
estimated mass of the SMBH located at Sgr~A* which results from this method
consistent with those of previous publications, but it also yields a good mass
estimate for the extragalactic source RE~J1034+396.

\def\arraystretch{1.3}
\begin{table}
\caption[Mass estimates from different methods for RE~J1034+396]{Mass estimates of the Seyfert~I 
galaxy RE~J1034+396 with different methods in chronological order.}
\begin{center}
\begin{tabular}{l l l}
    \hline
    \hline
      Publication & Mass & Method \\ \hline
      \citet{Gierlinski2008} & $6.3 \times 10^5 M_{\odot}$ & $H\beta$ \\
      \citet{Gierlinski2008} & $3.6 \times 10^7 M_{\odot}$ & $[O{III}]$ \\
      \citet{Gierlinski2008} & $(8 \times 10^6 - 9 \times 10^7) M_{\odot}$ & ISCO \\
      \citet{Bian2010} & $(1-4) \times 10^6 M_{\odot}$ & $M-\sigma_*$ \\
      \citet{Bian2010} & $(1-4) \times 10^6 M_{\odot}$ & $H\beta$ \\
      \citet{Jin2012} & $1.7 \times 10^6 M_{\odot} $ & $H\beta$ \\
    This paper & $1.421 \times 10^6 M_{\odot} $ & hotspot \\
    \hline
\end{tabular}
\end{center}\label{seyfert}
\end{table}

\section{Summary}
We have argued, that the double-peak structure of the X-ray flares observed from
Sgr~A* could arise from a simple orbiting hotspot model. In fact, only a fraction
of a full orbit is needed to result in a light curve with double-peak profile. It
is thus very probable that a hotspot can be stable long enough to create a
flare.

We have outlined a method which makes use of a comparison of the simulations
with the four brightest X-ray flares and gives an estimate on the mass of the
black hole. 
The mass estimate is independent of the uncertainties about the object distance.
The resulting masses are in close proximity to the other estimates of
the supermassive black hole mass which must be located at the position of the radio source Sgr~A*, which
make use of stellar orbits. Clearly the method should be tested by applying it to
other bright X-ray flares, observed in the future.
By applying this model to more and more flares, the estimate for the mass should
improve.
A future X-ray polarimeter mission would make this method even more reliable, 
because the polarization parameters can be simulated with this model as well.

The method described here works only is the light curves of the flare events are
dominated by the effects of relativistic motion of the luminous matter orbiting 
the black hole.
Hence, this method also has possible applications to other sources which exhibit flares
with a double-peak structure and could be used to get an estimate on the mass of
the black holes of sources, when stellar orbits cannot be resolved.
We also expect that the method works best on bright flares as these are
statistically less frequent. For more frequent, faint flares an overlap between flare
events is more likely and will therefore lead to less meaningful results.

\section{Acknowledgements}
We thank the referee for constructive comments and suggestions.
We received funding from the European Union Seventh Framework Program (FP7/2007-2013)
under grant agreement No. 312789 - Strong gravity: Probing Strong Gravity by Black
Holes Across the Range of Masses. This work was supported in part by the Deutsche
Forschungsgemeinschaft (DFG) via the Cologne Bonn Graduate School (BCGS), the Max
Planck Society through the International Max Planck Research School (IMPRS) for
Astronomy and Astrophysics, as well as special funds through the University of
Cologne and SFB 956 - Conditions  and Impact of Star Formation. M. Zajacek and
B. Shahzamanian are members of the IMPRS. Part of this work was supported by
fruitful discussions with members of the European Union funded COST Action MP0905:
Black Holes in a Violent Universe and the Czech Science Foundation - DFG collaboration
(No. 13-00070J).
We thank the German Ministry of Education and Research (BMBF) for
support under COPRAG2015 (No.57147386).

\bibliographystyle{mnras}
\bibliography{gc}

% Don't change these lines
\bsp	% typesetting comment
\label{lastpage}
\end{document}